# Toward light-controlled transverse Wigner ordering in attosecond electron beams

Yiming Pan[1†], Jiapeng Yang[1], Hao Geng[2,3††]

1. State Key Laboratory of Quantum Functional Materials, Center for Transformative Science and School of Physical Science and Technology, ShanghaiTech University, Shanghai 200031, China
2. College of Physics, Nanjing University of Aeronautics and Astronautics, Nanjing 211106, China
3. Key Laboratory of Aerospace Information Perception and Physics, MIIT, Nanjing 211106, China

**Abstract**

Kinetic fluctuations and space-charge expansion hinder Wigner ordering in free-electron beams. Here we propose a laser-driven route toward transient, substrate-free Wigner ordering that combines photon-induced near-field electron microscopy (PINEM) attosecond bunching, synchronized magnetic focusing, and spatial preconditioning. Classical relaxation and zero-temperature fixed-node diffusion Monte Carlo trap benchmarks show that sixfold order survives quantum fluctuations from 1 T to sub-Tesla fields, defining ultrafast preparation targets.

Wigner crystallization arises when Coulomb repulsion overwhelms kinetic energy and drives electrons toward an ordered charge configuration. Since Wigner's original proposal for a dilute electron gas [1], researchers have pursued this regime in confined platforms, such as quantum dots [2], electrons on helium [3], semiconductor heterostructures and high-mobility two-dimensional electron systems [4-8], and monolayer or moiré semiconductors [9–11]. These systems suppress kinetic fluctuations to reveal charge ordering, while their material environments shape the electronic state through confinement, screening, and disorder.

A freely propagating electron beam removes this material background but faces a different obstacle: the Coulomb interaction that favors ordering also drives rapid space-charge expansion. Finite emittance, transverse temperature, velocity spread, and partial incoherence further impede spatial ordering. Recent ultrafast electron microscopy studies have demonstrated strong Coulomb antibunching, attosecond phase-space correlations, and transient Coulomb crystallization in few-electron pulses [12–15]. The reported crystallization involves short axial chains in elongated, quasi-one-dimensional pulses. Recent theory also explores transverse guiding and continuous-beam crystallization [15]. Extending free-electron crystallization to the opposite, pancake-like quasi-two-dimensional regime, where tens of electrons could develop triangular correlations, requires controlling the areal density and lowering the effective transverse temperature to reach strong Coulomb coupling [16,17]. In conventional high-brightness beams, however, increasing the density strengthens space-charge forces, triggers disorder-induced heating as initially uncorrelated electrons relax [18], and shortens the time available for ordering. This competition between strong Coulomb coupling and beam stability is the central challenge for realizing free-electron Wigner crystallization.

Here, we propose a laser-driven route that combines photon-induced near-field electron microscopy (PINEM) [19–24], synchronized magnetic focusing, and spatial preconditioning. The PINEM interaction imprints an optical energy modulation that evolves during drift into a train of attosecond electron sheets [22], while magnetic focusing transiently increases the areal density. Spatial preconditioning could seed density modes compatible with triangular order and guide correlations during the short focal lifetime. By pre-aligning the electron density near low-potential-energy configurations before peak compression, spatial preconditioning could also limit disorder-induced heating [18] during conservative transport. Together, our longitudinal bunching analysis, classical energy minimization, and static-trap quantum Monte Carlo benchmarks define preparation targets and real- and reciprocal-space signatures of transient Wigner ordering. These benchmarks guide tests of whether the compressed beam can form and retain the target correlations.

### Optical Bunching and Magnetic Focusing

Figure 1 illustrates the proposed beamline for preparing a transient, strongly correlated electron sheet. A femtosecond-laser-triggered photocathode generates an ultrafast electron pulse with a controlled electron number, which we accelerate before it enters a nanostructured PINEM interaction region [19,20]. There, an optical near field imprints discrete energy sidebands on the longitudinal electron wavepacket; during subsequent drift, this modulation evolves into a train of attosecond-bunched electron sheets [21,22,25,26].

A synchronized magnetic condenser lens supplies transverse focusing. In the co-moving Larmor reference of a beam propagating along the $z$ axis, we represent an immersed axial magnetic field at the focal waist as an effective transverse harmonic trap and retain laboratory time. Magnetic focusing, residual transverse kinetic energy, and pairwise Coulomb repulsion govern the transverse dynamics.

We use a minimal Hamiltonian with longitudinal laser coupling, transverse magnetic confinement, and Coulomb
repulsion:

$$\mathcal{H} = \sum_{i=1}^{N}\left(\frac{\tilde{p}_{z,i}^2}{2\gamma^3 m} - ev_0 A(z_i,t) + \frac{\mathbf{p}_{\perp,i}^2}{2\gamma m} + \frac{1}{2}\gamma m \omega_\perp^2 r_i^2\right) + \frac{1}{\gamma}\sum_{i<j}^{N}\frac{e^2}{4\pi\epsilon_0|\mathbf{r}_i - \mathbf{r}_j|}. \quad (1)$$

Here, $N$ is the number of electrons, and $\boldsymbol{r}_i = (x_i, y_i)$ and $\boldsymbol{p}_{\perp,i}$ are the transverse coordinate and conjugate momentum of the $i$-th electron. We define $\tilde{p}_{z,i} = p_{z,i} - p_0$ relative to the central beam momentum $p_0 = \gamma m v_0$ in coordinates co-moving at velocity $v_0$. The effective longitudinal optical vector potential $A(z_i, t)$ (defined with the electron-charge sign absorbed) depends on coordinate $z_i$ and time $t$, and $\omega_\perp = eB/(2\gamma m)$ denotes the Larmor transverse trapping frequency for the axial magnetic field $B$ [27]. With electron rest mass $m$, charge $-e$, vacuum permittivity $\epsilon_0$, speed of light $c$, and Lorentz factor $\gamma = (1 - v_0^2/c^2)^{-1/2}$, the factors $\gamma m$ and $\gamma^3 m$ represent the relativistic transverse and longitudinal inertial masses, respectively. For coplanar electrons with equal longitudinal velocity, magnetic attraction partially cancels the laboratory electric repulsion, yielding the effective transverse Coulomb coupling coefficient $C = e^2/(4\pi\epsilon_0\gamma)$ [28].

Because the optical near-field interaction is shorter than the drift and focusing times, we separate the longitudinal evolution into an impulsive PINEM modulation stage and a subsequent free-drift stage. During the interaction stage, we neglect longitudinal drift and use $H_{\mathrm{I}} \approx -ev_0 \sum_{i=1}^{N} A\,(z_i, t)$. This impulsive coupling changes the phase and momentum spectrum of the electron wave packet without appreciably changing its instantaneous density (see Supplemental Material [29]). Figure 2a illustrates the PINEM phase-space pattern for film- or nanotip-based laser modulation [21,22]. After the packet exits the near field, we use the independent-electron drift Hamiltonian $H_{\mathrm{II}} = \sum_i \tilde{p}_{z,i}^2\,/(2\gamma^3 m)$. Free dispersion converts the PINEM sideband comb into Talbot-like temporal bunching at the compression plane [21–23], producing the attosecond electron sheets in Fig. 2b.

We synchronize the magnetic focus with the Talbot compression plane to minimize the transverse waist of the same attosecond-bunched sheet. This combined longitudinal compression and transverse focusing targets a quasi-two-dimensional regime where pairwise Coulomb repulsion competes with quantum zero-point motion and residual thermal agitation. In this Larmor reference, we quantify the competition between transverse Coulomb repulsion and harmonic magnetic confinement with the dimensionless quantum coupling parameter

$$\kappa(B) = \frac{C}{\hbar\omega_\perp l_q} = \frac{l_q}{a_0} = \sqrt{\frac{2\hbar}{eBa_0^2}} = \frac{\kappa_0}{\sqrt{B/(1\,\mathrm{T})}}, \tag{2}$$

We define this parameter as the ratio of the characteristic interaction energy $C/l_q$ at the transverse harmonic-oscillator zero-point length $l_q = \sqrt{\hbar/(\gamma m\omega_\perp)} = \sqrt{2\hbar/(eB)}$ to the oscillator energy quantum $\hbar\omega_\perp$ (with $\hbar$ the reduced Planck constant). Because the Lorentz factor $\gamma$ cancels between the transverse mass $\gamma m$ and the Coulomb coefficient $C = e^2/(4\pi\epsilon_0\gamma)$, $\kappa(B)$ reduces to the ratio of $l_q$ to the Bohr radius $a_0 = 4\pi\epsilon_0\hbar^2/(me^2) \simeq 0.0529\,\mathrm{nm}$. Here $\kappa_0 \equiv \sqrt{2\hbar/[e(1\,\mathrm{T})a_0^2]} \simeq 686$ denotes the characteristic quantum coupling constant at a unit reference focusing field of $B = 1\,\mathrm{T}$, fixed solely by fundamental constants ($l_q(1\,\mathrm{T}) \simeq 36.3\,\mathrm{nm}$). Reducing the reference field lowers the zero-point kinetic energy $\hbar\omega_\perp \propto B$ faster than the characteristic Coulomb energy $C/l_q \propto B^{1/2}$, analogous to lowering the carrier density (increasing the Wigner–Seitz radius $r_s$) in a two-dimensional electron gas. In this static model, reducing $B$ from 1 T into the sub-Tesla regime (0.1–0.5 T) expands the electron sheet while increasing $\kappa(B)$ into the $(1\text{–}2) \times 10^3$ range, where Coulomb repulsion strongly dominates over quantum zero-point motion.

At large $\kappa$, thermal agitation imposes an additional preparation constraint: the characteristic Coulomb energy $E_C = C/a_{\mathrm{eq}}$ (evaluated at the effective Wigner–Seitz radius $a_{\mathrm{eq}}$) must exceed the residual transverse thermal energy $k_B T_\perp$ (with $k_B$ the Boltzmann constant and $T_\perp$ the effective transverse temperature). We use the classical coupling ratio $\Gamma = E_C/(k_B T_\perp)$ of order $10^2$ as a strong-coupling preparation target [16,17]. Representative optical settings and Larmor-frame preparation targets for an attobunched 200 keV electron sheet ($N \sim 10\text{–}10^2$, driven at $\lambda_L = 800\,\mathrm{nm}$ with phase-modulation amplitude $g = 6\text{–}10$) are detailed in the Supplemental Material [29]. Comparing the 1 T reference regime ($\kappa = \kappa_0$, Fig. 3) with the sub-Tesla regime ($B = 0.1\text{–}0.5\,\mathrm{T}$) shows how weaker focusing enlarges the target sheet ($a_{\mathrm{eq}} \simeq 0.29\text{–}0.86\,\mu\mathrm{m}$) and relaxes the geometric bunch-duration ($\tau_b \simeq 100\text{–}300\,\mathrm{as}$) and normalized thermal-emittance ($\varepsilon_n \lesssim 11\text{–}18\,\mathrm{pm\,rad}$) budgets, while requiring a lower $T_\perp \lesssim 0.14\text{–}0.41\,\mathrm{K}$ [29].

Because unconfined axial expansion doubles the thin-sheet thickness within $\sim 0.3\,\tau_C$ [29], we propose spatial preconditioning before maximum compression to shorten the time required for ordering. Tailoring the transverse focusing profile or using a structured PINEM modulation could seed density modes near the reciprocal-lattice vectors of triangular Wigner order. Such seeds could steer the early dynamics without imposing a rigid lattice. During the brief focal window, the electron sheet may develop triangular spatial correlations, sixfold structure-factor peaks, and enhanced bond-orientational order, consistent with a Wigner-ordered configuration. Here we use crystallization to denote a finite, nonequilibrium, transient Coulomb-dominated ordered state targeted through synchronized longitudinal compression and transverse focusing.

**Transient Crystallization and Quantum Correlations**

We synchronize the magnetic focal waist with the PINEM Talbot compression plane to target a thin microbunch sheet whose co-moving longitudinal thickness ($\gamma\Delta z \sim 0.1\, a_{\mathrm{eq}}$) is an order of magnitude smaller than the effective Wigner–Seitz radius $a_{\mathrm{eq}}$. Assuming upstream single-bunch gating (or compensation of smooth inter-sheet fields in the net focusing strength [29]), within this target thin-sheet window before axial space-charge rebound, we approximate pairwise Coulomb repulsion by the coplanar $1/r_{\perp}$ interaction in the Larmor reference [29]. Because conservative focusing compresses the spatial envelope without dissipative cooling, we examine transverse Wigner ordering with two complementary benchmarks: the classical zero-temperature point-charge limit ($\hbar \to 0, T_{\perp} \to 0$, Figs. 2c,d) to identify finite-cluster packing, and zero-temperature quantum Monte Carlo simulations ($\hbar \neq 0$, Fig. 3) [29] to test whether sixfold order survives quantum zero-point motion and Pauli exchange.

In the classical point-charge limit, we determine ordered configurations by minimizing the potential energy of repelling charges in the transverse trap. Neutralized bulk-lattice studies favor triangular coordination [32] and provide a local packing reference. The finite trap has no neutralizing background; its confinement-plus-repulsion energy, shell occupation, and boundary determine the complete classical configuration [17,33].

We use stochastic gradient relaxation to minimize the transverse confinement-plus-repulsion potential [29]. Here the relaxation steps represent a fictitious numerical quench schedule used to locate low-energy static configurations, rather than the real-time magnetic-focusing duration or physical beam cooling. Starting from a disordered two-dimensional electron cloud [Fig. 2(c)], numerical annealing, with coordinates in units of the classical balance length $l_{\mathrm{cl}} = [C/(\gamma m \omega_{\perp}^{2})]^{1/3}$, reduces fivefold and sevenfold disclination defects and yields a low-energy packing motif [Fig. 2(d)] [29]: concentric outer shells shaped by harmonic confinement around an interior with sixfold-coordinated triangular Wigner order.

Classical relaxation [Fig. 2(d)] treats electrons as point charges; we next test whether sixfold order survives quantum zero-point motion in the zero-temperature limit ($T_{\perp} \to 0$). The quantum coupling $\kappa(B)$ [Eq. (2)] is independent of $N$, while harmonic confinement–repulsion balance yields only weak $N^{\pm 1/6}$ variations in effective Wigner–Seitz radius $a_{\mathrm{eq}}$ and Coulomb scale $E_C$ across $N \sim 10$–$10^{2}$ [cf. $N = 120$ in Fig. 2(d)] [29]. We benchmark this zero-temperature limit for a representative spin-polarized closed-shell cluster in the static transverse trap using variational and fixed-node diffusion Monte Carlo (VMC and FN-DMC [29,33–36]). At fixed electron number and spin polarization, scaling lengths by the oscillator width $l_q$ leaves $\kappa(B)$ as the only continuous coupling parameter. It sets the ratio $l_{\mathrm{cl}}/l_q = \kappa^{1/3}$ between the classical balance length and the single-particle zero-point width ($l_{\mathrm{cl}}/l_q \simeq 8.8$ at $B = 1\,\mathrm{T}$). In Fig. 3(d), the FN-DMC energy estimate per electron $E_{\mathrm{DMC}}/N$ (blue circles) lies systematically below the Jastrow-free scaled-Slater upper bound $E_{\mathrm{ub}}/N$ (black dashed curve, retaining the exact harmonic center-of-mass width [29]) across $\kappa \geq 1$. The correlation-energy lowering $\Delta E_{\mathrm{corr}}/N \equiv (E_{\mathrm{ub}} - E_{\mathrm{DMC}})/N$ [Fig. 3(d), inset] approaches the $\kappa^{2/3}$ scaling at strong coupling, quantifying the energy gained beyond this Jastrow-free reference. Comparing $E_{\mathrm{ub}}$ with FN-DMC energies tests amplitude correlations within a common nodal family, rather than locating a liquid–crystal crossing or quantum melting threshold.

Figures 3(a)–3(c) show real- and reciprocal-space correlations of this FN-DMC state at $B = 1\,\mathrm{T}$ ($\kappa = \kappa_0 \simeq 686$). The circular trap has no preferred azimuth. We align each sampled configuration by its interior sixfold bond angle before averaging [29]. This alignment reveals concentric shells surrounding a localized triangular core in the aligned one-body density $n(\boldsymbol{r})$ [Fig. 3(a)] and six sharp Bragg-like peaks along the primary reciprocal ring of the aligned transverse static structure factor $S(\boldsymbol{q})$ [Fig. 3(b)]. The outermost ring in Fig. 3(a) is smooth under inner-core alignment. We place an outer-shell reference electron on the $+x$ axis (cyan marker) by rotating sampled coordinates. The co-rotating conditional density $n_{\mathrm{cond}}(\boldsymbol{r})$ [Fig. 3(c)] then reveals an exchange-correlation hole flanked by sharp nearest-neighbor peaks along the outer ring [37,38]. From weak coupling ($\kappa \leq 1$) to 1 T and into softer sub-Tesla traps ($B = 0.5$–$0.1\,\mathrm{T}$), the interior sixfold bond-orientational order parameter $|\psi_6|$ rises more than fivefold from $0.12$ to $0.66$–$0.74$ [29].

We must reconstruct the finite, short-lived state's spatial correlations over nominally identical shots. Comparing the real-space patterns in Figs. 2(d) and 3(a) with detected signals requires a propagation and detector-response model. To extract $S(\boldsymbol{q})$ from particle imaging, we must record multi-electron arrival positions within individual microbunches, for example through high-magnification conjugate imaging of the focal waist with a target object-plane resolution of $\lesssim 30$–$70\,\mathrm{nm}$. Ultrafast diffraction offers a possible alternative; either route requires a signal model, and averaging single-particle density images alone cannot recover the full structure factor. We propose detuning the PINEM laser timing or magnetic lens current as experimental controls to test whether sixfold reciprocal-space peaks weaken when spatiotemporal compression is lost.

The static-trap correlations motivate further tests of the strong-coupling origin and dynamical survival of ordering. At fixed focusing conditions, an excessive electron number ($N$) can enhance space-charge expansion and disrupt local order, while greater transverse kinetic spread competes with relative localization; longitudinal energy spread affects the attainable microbunch width. Time-dependent beam simulations would test whether a finite-size, nonequilibrium Wigner-ordered configuration can form near the focus and determine how long it could survive before expansion and kinetic fluctuations disrupt it. The target observable is a transient structure factor consistent with triangular electron correlations.

We therefore target a transient, finite-size, freely propagating charge-ordered configuration. The proposal applies Coulomb-crystal and non-neutral-plasma concepts to a laser-compressed electron beam without a substrate or static lattice. The sought evidence is the emergence of triangular correlations when Coulomb repulsion dominates the residual kinetic energy in the attosecond sheet. Ultrafast optical modulation and magnetic beam control thus define a route toward testing material-free Wigner ordering.

## Further Discussion

The proposed scheme targets a transient, substrate-free strongly correlated state in a freely propagating electron beam. PINEM modulation, magnetic focusing, beam energy, and energy spread provide controls over the sheet density, pulse duration, and effective interaction strength. This tunability could make the finite electron sheet a platform for probing how Coulomb order forms, melts, and supports collective modes on ultrafast time scales. This perspective also extends quantum electron optics [24–26,39,40] beyond single-particle wave-packet shaping. Attosecond longitudinal bunching prepares the temporal beam structure. Our transverse trap benchmarks identify the spatial correlations to target near the magnetic focus. The proposal thus links PINEM-based coherent control to collective electron ordering, suggesting a route toward engineered two-dimensional electron configurations and structured attosecond bunches.

Several experimental and theoretical challenges remain. Observing transient crystallization would require high-efficiency position-sensitive detection, shot-to-shot reproducibility, and synchronization of the PINEM compression with the magnetic focus. Emittance, energy spread, electron number, and space-charge expansion constrain the operating window; cold-beam dynamics can also produce density caustics [41], so experiments must distinguish caustics from interparticle order. The broader context includes trapped nonneutral-plasma and charged-particle Coulomb crystals [42], but the freely propagating, light-bunched geometry considered here will require time-dependent many-body quantum beam simulations beyond equilibrium trap models. Future work should connect the classical relaxation and QMC benchmarks to beam dynamics and quantify how preconditioning, quantum delocalization, exchange, and spin preparation affect transient ordering.

## Conclusion

We propose a laser-driven route toward transient Wigner order in a freely propagating electron beam. We combine PINEM-induced attosecond bunching with synchronized transverse magnetic focusing to target a strongly coupled quasi-two-dimensional electron sheet. Our classical relaxation and static-trap quantum Monte Carlo benchmarks show pronounced sixfold orientational order and shell localization, and these signatures persist in the fixed-node calculations from 1 T down to sub-Tesla reference fields. The static-trap FN-DMC results do not by themselves establish formation or survival of order in the propagating beam, and thus time-dependent calculations must test both within the transient focal window. These benchmarks define target real-and reciprocal-space correlation signatures and provide quantitative phase-space targets for future time-dependent beam simulations and ultrafast diffraction experiments.

Our work suggests a route toward material-free strongly correlated electron matter controlled by light and magnetic optics. The proposed platform could enable ultrafast studies of disorder-to-order dynamics, melting, and collective modes in finite Coulomb systems. It also connects attosecond free-electron beam engineering with many-body correlations, complementing proposals for light-mediated electron pairing [40], with possible implications for quantum electron optics [39], beam physics, and coherent radiation from structured electron bunches.

**Acknowledgments**

Y.P. acknowledges support from the NSFC (No. 2023X0201-417-03) and the start-up fund from ShanghaiTech University. H.G. acknowledges support from the National Natural Science Foundation of China (Grant No. 12304068), the Startup Fund of Nanjing University of Aeronautics and Astronautics (Grant No. YAH24076), and the Fundamental Research Funds for the Central Universities (Grant No. NS2026035).

Address correspondence and requests for materials to Y.P. (†yiming.pan@shanghaitech.edu.cn) and H.G. (††genghao@nuaa.edu.cn).

**Data and code availability**

We store the simulation code and calculation parameters in a GitHub repository: https://github.com/genghaophy/FreeElectronWignerCrystal. We will make the repository public at or after publication.

**Conflict of interest**

The authors declare no competing interests.

**Figures:**

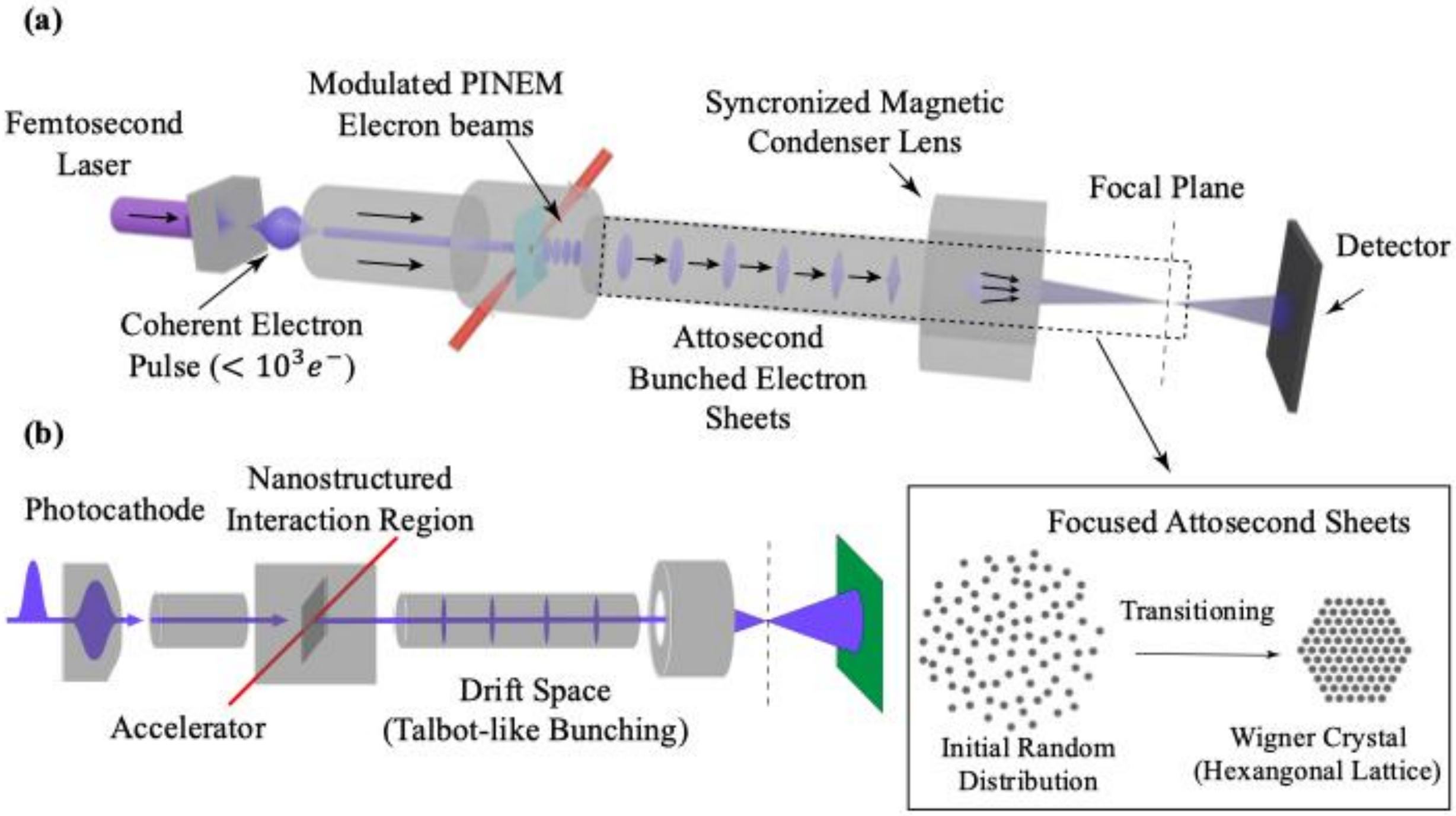

**Figure 1:** Ultrafast free-electron route toward transient Wigner ordering. (a) Laser-driven beamline. A laser-triggered photocathode emits a packet with $N < 10^3$ electrons before microbunch selection. After acceleration, PINEM imprints energy sidebands; free drift converts them into attosecond bunches. A synchronized magnetic lens focuses the beam transversely at the longitudinal compression plane, targeting a dense, quasi-two-dimensional sheet. (b) Schematic longitudinal evolution from emission to detection. The ordered inset illustrates the target triangular correlations.

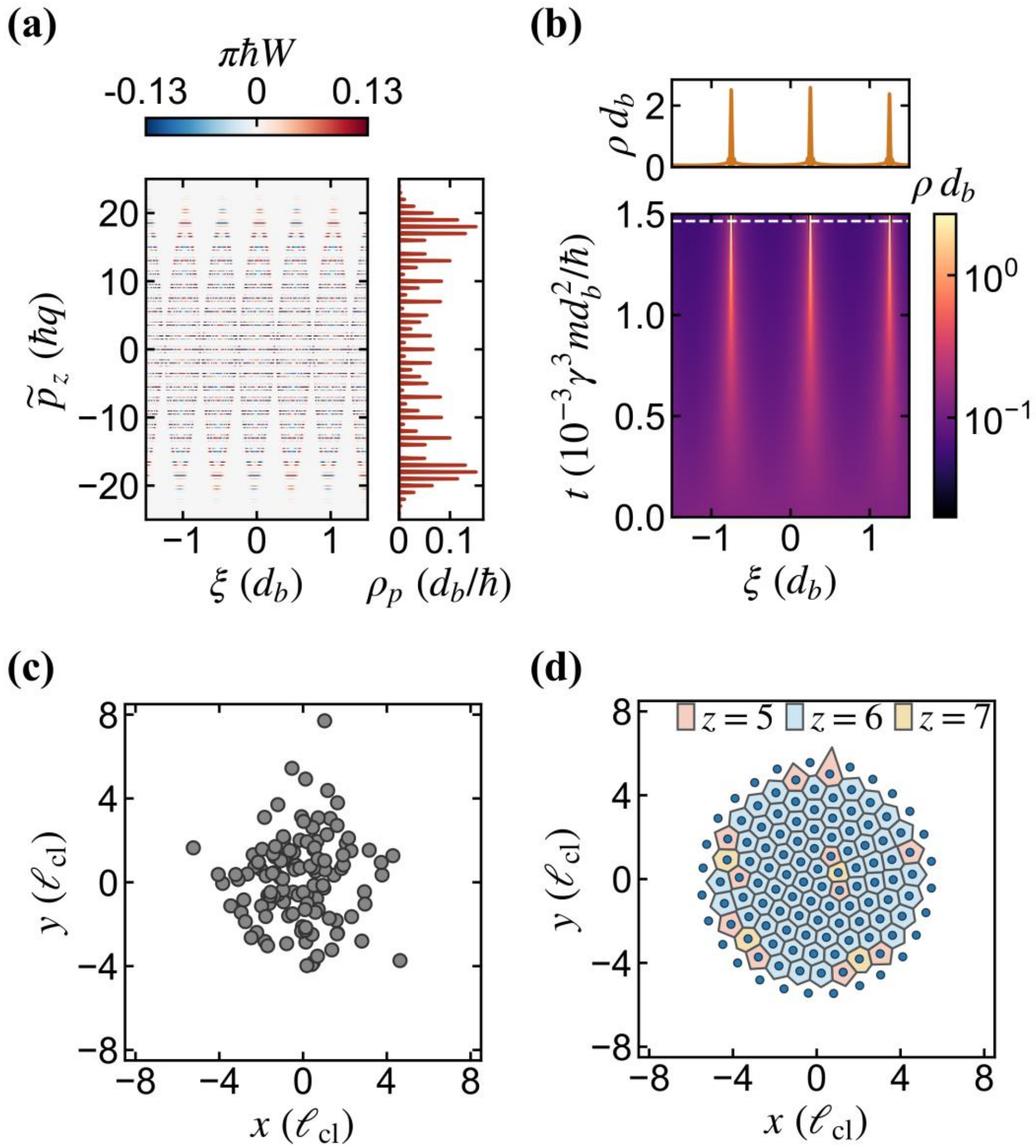

**Figure 2:** Longitudinal attobunching and classical transverse ordering. (a) PINEM-modulated Wigner function $W(\tilde{z}, \tilde{p}_z)$ for modulation strength $g = 20$, initial envelope width $\sigma_\xi = 3d_b$, and drift time $t_* = \hbar t/(\gamma^3 m d_b^2) = 3 \times 10^{-4}$ (colors show $\pi\hbar W$ clipped to $\pm 0.13$ to resolve interference fringes). The axes show $\xi \equiv \tilde{z}$ in units of $d_b = 2\pi/q$ and $\tilde{p}_z$ in units of $\hbar q$; the right panel shows the momentum marginal $\rho_p$. (b) Independent-electron drift produces attosecond density peaks in $\rho(\xi, t)$; the upper trace shows the density at the dashed line ($t_* \simeq 1.47 \times 10^{-3}$, near the $1.55 \times 10^{-3}$ focus). (c,d) Separate classical benchmark for $N = 120$ point charges in a harmonic trap, before (c) and after (d) numerical quench minimization (fictitious annealing steps rather than physical focusing time). Transverse lengths are in units of $l_{\rm cl}$. The relaxed configuration has a triangular core and outer rings; Voronoi colors mark coordination numbers $z = 5,6,7$.

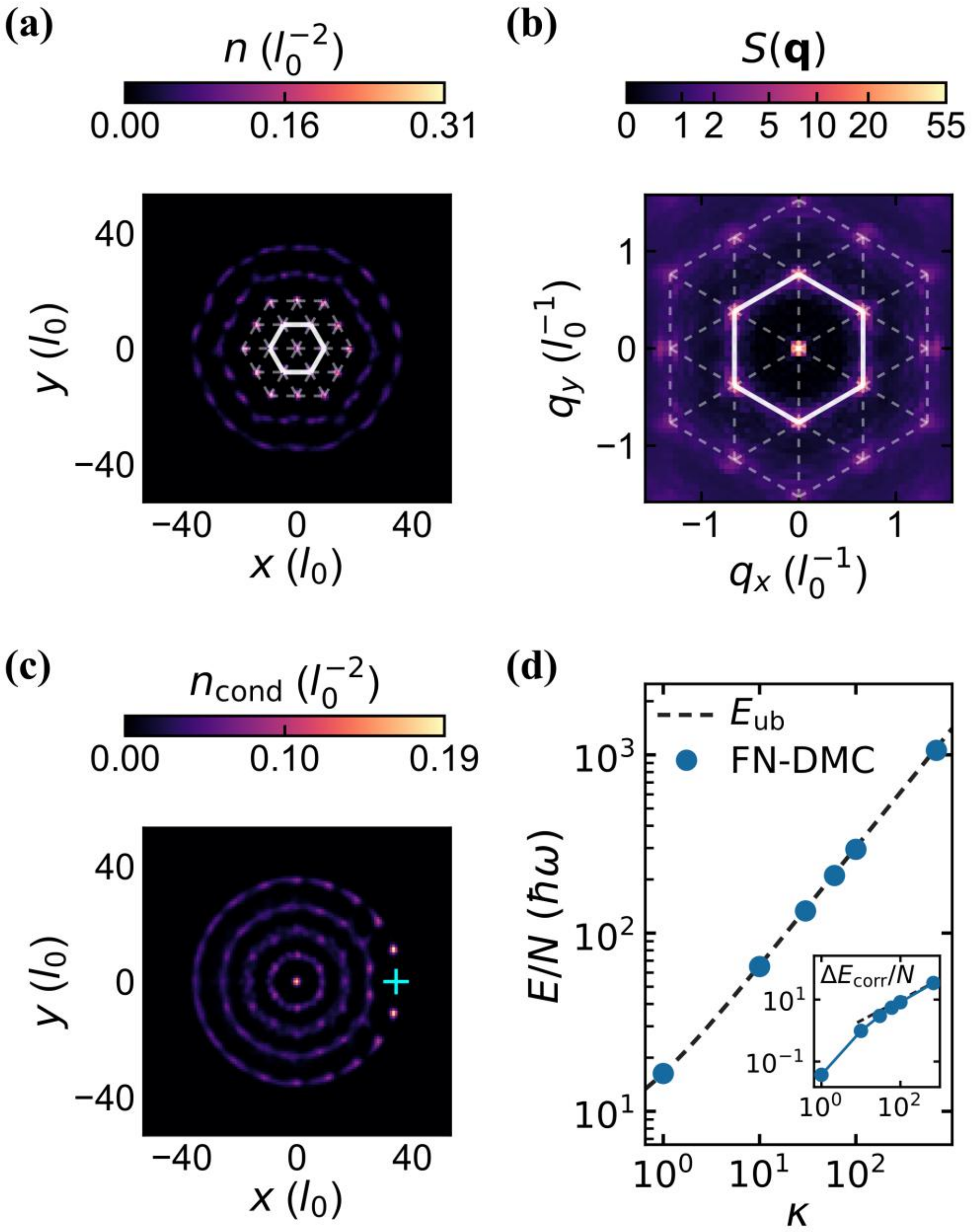

**Figure 3:** Zero-temperature quantum correlations in a 2D harmonic trap with $1/r$ Coulomb repulsion. FN-DMC treats $N = 55$ fully spin-polarized electrons; lengths are in units of $l_q$ (labeled $l_0$). Panels (a–c) show results at 1 T ($\kappa = \kappa_0 \simeq 686$). (a) Density $n(\boldsymbol{r})$, aligned by the inner-core sixfold bond angle, resolves a triangular core and outer shells. (b) Structure factor $S(\boldsymbol{q})$ in the same frame shows six reciprocal-space peaks; colors map $\ln[1 + S(\boldsymbol{q})]$ and ticks give $S(\boldsymbol{q})$. White lines are geometric guides. (c) Conditional density $n_{\mathrm{cond}}(\boldsymbol{r})$, rotated to place an outer-shell reference electron at $+x$ (cyan), reveals an exchange-correlation hole and neighboring peaks [37,38]. (d) Energy per electron $E/N$ in units of $\hbar\omega_\perp$ (labeled $\hbar\omega$) versus $\kappa$: FN-DMC $E_{\mathrm{DMC}}/N$ (blue) and the Jastrow-free scaled-Slater upper bound $E_{\mathrm{ub}}/N$ (black dashed). Inset: energy lowering $\Delta E_{\mathrm{corr}}/N \equiv (E_{\mathrm{ub}} - E_{\mathrm{DMC}})/N$ in the same units versus $\kappa$ over the same coupling range; the dashed guide shows asymptotic strong-coupling $\propto \kappa^{2/3}$ scaling.

# Supplemental Material

# Toward light-controlled transverse Wigner ordering in attosecond electron beams

Yiming Pan[1†], Jiapeng Yang[1], Hao Geng[2,3††]

4. State Key Laboratory of Quantum Functional Materials, Center for Transformative Science and School of Physical Science and Technology, ShanghaiTech University, Shanghai 200031, China
5. College of Physics, Nanjing University of Aeronautics and Astronautics, Nanjing 211106, China
6. Key Laboratory of Aerospace Information Perception and Physics, MIIT, Nanjing 211106, China

## Section A: Analytical Formulation of PINEM and Attobunching

**Derivation of the Hamiltonian from Relativistic Kinematics**. For an electron interacting with an optical near field, we expand the relativistic energy–momentum relation $E = \sqrt{(p_z - eA_z)^2c^2 + m^2c^4}$. For a bunch with central momentum $p_0 = \gamma m v_0$, we assume that the momentum fluctuations $\delta p_z = p_z - p_0$ and near-field momentum kick $eA_z$ are small compared with $p_0$ and Taylor-expand the dispersion in the kinetic momentum $\Pi_z = p_z - eA_z$ about $p_0$ (using $\Pi_z = p_z - eA_z$, with the sign of the electron charge $-e$ absorbed into the convention for $A_z$).

The unperturbed energy is $E(p_0) = \gamma mc^2$. The first derivative evaluated at $p_0$ returns the classical bunch velocity $v_0$. The second derivative defines the well-known longitudinal mass $m_L$:

$$\left.\frac{\partial^2 E}{\partial \Pi_z^2}\right|_{p_0} = \frac{1}{\gamma m}\left(1 - \frac{v_0^2}{c^2}\right) = \frac{1}{\gamma^3 m} \equiv \frac{1}{m_L}$$

Thus, expanding to second order, we obtain:

$$E(p_z) \approx \gamma mc^2 + v_0(p_z - p_0 - eA_z) + \frac{(p_z - p_0 - eA_z)^2}{2\gamma^3 m} \qquad (S1)$$

A time-dependent canonical transformation to co-moving longitudinal coordinates introduces the relative momentum $\tilde{p}_z = p_z - p_0$ and, after constant energy offsets are dropped, gives the effective longitudinal Hamiltonian:

$$\tilde{H}_z = -ev_0A_z + \frac{(\tilde{p}_z - eA_z)^2}{2\gamma^3 m} \qquad (S2)$$

Whenever the momentum spread and near-field kick remain small compared with $p_0$ ($|\tilde{p}_z|, |eA_z| \ll p_0$) and longitudinal dispersion is negligible across the sub-micrometer interaction zone, we retain the linear coupling $-ev_0A_z$ during the impulsive PINEM stage and the quadratic dispersion term $\tilde{p}_z^2/(2\gamma^3 m)$ during subsequent drift. This separation motivates the two-stage treatment used in the main text.

Within the independent-electron approximation, we solve the time-dependent Schrödinger equation analytically for both stages. Let the initial longitudinal wave packet entering the interaction region be a broad, smooth function $\psi_0(\tilde{z})$ of the co-moving longitudinal coordinate $\tilde{z} = z - v_0 t$.

**Stage I: Quantum Phase Modulation.** Over the short interaction length, spatial drift is neglected. We retain only the linear PINEM interaction $\mathcal{H}_I = -ev_0 A(\tilde{z}, t)$ in this impulsive approximation. Representing the sampled optical near field by the effective quasi-stationary vector-potential grating $A(\tilde{z}, t) = A_0 \sin(q\tilde{z} + \phi_0)$ in the co-moving coordinate, we obtain the electron's eikonal phase by integrating over the interaction time:

$$\psi_I(\tilde{z}) = \psi_0(\tilde{z}) \exp[ig \sin(q\tilde{z} + \phi)] \qquad (S3)$$

where $A_0$ and $\phi_0$ are the peak amplitude and initial phase of the effective vector-potential grating, $g$ is the dimensionless phase-modulation amplitude proportional to the interaction length and field amplitude, $\phi$ is the accumulated phase offset (set to $\phi = 0$ below by choosing the co-moving coordinate origin), and $q = \omega_L/v_0$ is the effective longitudinal modulation wave number, with $\omega_L$ the laser angular frequency. Applying the Jacobi-Anger identity, we decompose the pure sinusoidal phase modulation into a coherent superposition of discrete momentum sidebands:

$$\psi_I(\tilde{z}) = \psi_0(\tilde{z}) \sum_{n=-\infty}^{\infty} J_n(g) e^{inq\tilde{z}} \qquad (S4)$$

where $J_n(g)$ represents the $n$-th order Bessel function of the first kind. Within this phase-only approximation, the real-space density $|\psi_I(\tilde{z})|^2 = |\psi_0(\tilde{z})|^2$ remains unchanged, while the momentum distribution acquires photon sidebands.

**Stage II: Temporal Talbot Effect and Bunching.** In the subsequent drift region, the optical field vanishes ($A_0 = 0$). The longitudinal Hamiltonian is then $\mathcal{H}_{II} = \frac{\tilde{p}_z^2}{2\gamma^3 m}$. Over a free-flight time $t_d$, the propagator adds a quadratic phase to each plane-wave component $k_n = nq$:

$$\psi_{\text{final}}(\tilde{z}, t_d) = \psi_0(\tilde{z}) \sum_{n=-\infty}^{\infty} J_n(g) e^{inq\tilde{z}} \exp\left(-i \frac{\hbar q^2 t_d}{2\gamma^3 m} n^2\right) \qquad (S5)$$

The exponential phase factor $\exp(-i\theta n^2)$, governed by the dimensionless sideband dispersion parameter $\theta = \frac{\hbar q^2 t_d}{2\gamma^3 m}$, generates temporal Talbot evolution. Equation (S5) uses the broad-envelope approximation ($\sigma_{\tilde{z}} \gg d_b = 2\pi/q$) and neglects envelope evolution during drift, whereas Figs. 2(a,b) propagate the full modulated wave packet in Fourier space with the quadratic free-space propagator $\exp[-i\hbar k^2 t_d/(2\gamma^3 m)]$; with length normalized by the optical period $d_b = 2\pi/q$, $\theta$ relates to the dimensionless drift time $t_* = \hbar t_d/(\gamma^3 m d_b^2)$ in Figs. 2(a,b) through $\theta = (q d_b)^2 t_*/2 = 2\pi^2 t_*$. For $\theta \ll 1$, differential sideband phases convert the momentum comb into longitudinal density modulation; using the velocity difference $\Delta v_n \approx \frac{n\hbar q}{\gamma^3 m}$, the ray model predicts a caustic at $\theta_c = 1/(2g)$ ( $t_{*,c} = [4\pi^2 g]^{-1} \simeq 1.27 \times 10^{-3}$ for $g = 20$ ), while the wave calculation gives a density maximum slightly downstream ($t_{*,\max} \simeq 1.55 \times 10^{-3}$).

At the compression plane, sideband interference produces peaks in the longitudinal density $\rho(\tilde{z}) = |\psi_{\text{final}}|^2$, corresponding to a train of attosecond-scale electron sheets separated in arrival time by the optical period. This compression provides the thin-sheet initial geometry assumed in the transverse ordering calculation.

**Illustrative Optical Parameters and Short-Wavelength Inter-Sheet Coupling.** To connect Eqs. (S3)–(S5) with representative ultrafast electron-microscopy conditions (distinct from the strong-modulation $g = 20$ phase-space example in Fig. 2), we consider an illustrative 200 keV electron beam ($\gamma \simeq 1.39$, $v_0 \simeq 0.70c$) modulated at an optical wavelength $\lambda_L = 800$ nm ($\omega_L = 2\pi c/\lambda_L$, $\hbar\omega_L \simeq 1.55$ eV ) to prepare microbunch sheets with target population $N \sim 10$–$10^2$ . A dimensionless PINEM phase-modulation amplitude $g = 6$–$10$ imprints a peak longitudinal energy modulation $g\hbar\omega_L \simeq 9$–$16$ eV. Accumulated over an illustrative phase-matched near-field path $L_{\text{phase}} \simeq 0.1\ \mu$m, this modulation corresponds to an effective peak longitudinal electric field $E_{\text{pk}} = g\hbar\omega_L/(eL_{\text{phase}}) \simeq 0.12$ V/nm for $g = 8$ and a vacuum-equivalent cycle-averaged intensity $I_{\text{eq}} = \epsilon_0 c E_{\text{pk}}^2/2 \simeq 2$ GW cm$^{-2}$ (the required incident laser intensity depends on the specific near-field enhancement geometry). In the zero-energy-spread limit of Eq. (S5), subsequent free-space drift places the ray caustic at $L_c \simeq 3.8$–$6.3$ mm and the first wave-optical focus at $L_d \simeq 5.0$–$9.0$ mm across $g = 10$–$6$, concentrating 45%–47% of the electrons per optical cycle within a 150 as window around a central spike of 90–144 as full width at half maximum (for $g = 8$ at $L_d \simeq 6.5$ mm, 47% of the cycle charge falls within 150 as at 110 as FWHM; cf. background-subtracted experimental pulse trains in Refs. [S1,S2]). Assuming temporal gating isolates this central spike from the unbunched cycle pedestal, combining longitudinal compression with an immersed sub-Tesla magnetic focus ($B = 0.1$–$0.5$ T; Section B) enlarges the transverse cluster and relaxes the geometric thin-sheet target ($\gamma v_0 \tau_b / a_{\text{eq}} = 0.1$) to selected-sheet rms durations $\tau_b \simeq 100$–$300$ as. Successive 800 nm microbunches are separated by $D' = \gamma v_0 T_L \simeq 0.77\ \mu$m in the mean-electron rest frame ($T_L = \lambda_L/c$). Without upstream single-bunch gating, neighboring sheets contribute a smooth radial defocusing field comparable in magnitude to $\gamma m \omega_\perp^2 r$ when $D' < 2\sigma_x$ (which must be compensated in the net focusing strength or removed by upstream gating), whereas the short-wavelength inter-sheet Coulomb component at the first triangular reciprocal-lattice vector $G_1 = 4\pi/(\sqrt{3}d_{\text{nn}})$ (with nearest-neighbor spacing $d_{\text{nn}} = (2\pi/\sqrt{3})^{1/2}a_{\text{eq}} \simeq 1.90\, a_{\text{eq}}$) is exponentially suppressed as $e^{-G_1 D'} \simeq 3 \times 10^{-2}$ to $4 \times 10^{-5}$ across $B = 0.1$–$0.5$ T.

**Section B: Classical and Quantum Treatments of Wigner Crystallization**

Section A models independent-electron PINEM modulation and Talbot compression toward selected microbunches of co-moving thickness $\gamma\Delta z \ll a_{\text{eq}}$; here we benchmark an ideal two-dimensional transverse many-body model for a synchronized magnetic focus immersed in an axial field $B$. At the compression waist ($\gamma\Delta z \sim 0.1\, a_{\text{eq}}$), we approximate the instantaneous pairwise Coulomb potential by the coplanar $1/r_\perp$ form, with characteristic relative corrections $\mathcal{O}[(\gamma\Delta z/a_{\text{eq}})^2] \sim 10^{-2}$ at separations near the characteristic interelectron spacing, in the co-moving Larmor frame prior to axial space-charge rebound. First, in the classical zero-temperature point-charge limit ($\hbar \to 0$, $T_\perp \to 0$), rescaling transverse coordinates by the classical confinement–repulsion balance length $l_{\text{cl}} \equiv l_0$ isolates the universal geometric packing landscape $U$ [Eq. (S7)], which we relax via a numerical stochastic quench [Eq. (S10)] to obtain a low-energy shell-plus-triangular motif (Figs. 2c,d). Second, in the quantum many-body regime ($\hbar \neq 0$), rescaling by the harmonic zero-point length $l_q = \kappa^{-1/3} l_{\text{cl}}$ expresses the Hamiltonian at fixed particle number in terms of the dimensionless coupling $\kappa(B)$ [Eqs. (S8)–(S9)], and variational and fixed-node diffusion Monte Carlo (VMC and FN-DMC) simulations [Eqs. (S11)–(S13)] assess crystalline correlations within the selected spin sector and fixed-node approximation (Fig. 3).

**Transverse Hamiltonian and Dimensionless Scaling**. Near the longitudinal compression waist within an immersed axial magnetic field $B$, we model an isolated attosecond sheet in the co-moving Larmor $(x, y)$ plane with an effective harmonic restoring potential of frequency $\omega_\perp = eB/(2\gamma m)$. We include transverse kinetic energy, harmonic confinement, and coplanar Coulomb repulsion:

$$\mathcal{H}_\perp = \sum_{i=1}^{N} \left[ \frac{\mathbf{p}_{\perp,i}^2}{2\gamma m} + \frac{1}{2}\gamma m \omega_\perp^2 \mathbf{r}_{\perp,i}^2 \right] + \frac{1}{\gamma} \sum_{i<j}^{N} \frac{1}{4\pi\epsilon_0} \frac{e^2}{|\mathbf{r}_{\perp,i} - \mathbf{r}_{\perp,j}|} \qquad (S6)$$

Here, $\boldsymbol{r}_{\perp,i}$ and $\boldsymbol{p}_{\perp,i}$ are the two-dimensional transverse coordinate and momentum vectors. Magnetic attraction $F_B$ partially cancels the laboratory transverse electric repulsion $F_E$, giving a net force $F_{\text{net}} = F_E - F_B = F_E/\gamma^2$; combined with the Lorentz enhancement of the transverse electric field for coplanar co-moving electrons, this yields the effective Coulomb pair coefficient $C = \frac{e^2}{4\pi\epsilon_0 \gamma}$ [S3]. The transverse inertial mass is $\gamma m$, distinct from the longitudinal mass $\gamma^3 m$ that controls Talbot-like bunching.

To identify the classical ordering motif in the zero-temperature point-charge limit ($\hbar \to 0$, $\mathbf{p}_{\perp,i} = 0$), we first isolate the potential-energy landscape. We introduce the classical characteristic length $l_0 = [C/(\gamma m\omega_\perp^2)]^{1/3} = \left(\frac{e^2}{4\pi\epsilon_0\gamma^2 m\omega_\perp^2}\right)^{1/3}$ (denoted $\ell_{\mathrm{cl}}$ in Fig. 2), defined by equality of the harmonic confinement scale $\gamma m\omega_\perp^2 l_0^2$ and the Coulomb scale $C/l_0$ ; with dimensionless coordinates $\tilde{\mathbf{r}}_i = \mathbf{r}_{\perp,i}/l_0$ and classical energy unit $E_{\mathrm{cl}} = \gamma m\omega_\perp^2 l_0^2 = C/l_0$, the physical beam and lens parameters $(\gamma, B, \omega_\perp)$ drop out at fixed particle number, giving the universal form:

$$U(\{\tilde{\boldsymbol{r}}_i\}) = \sum_{i=1}^{N}\frac{1}{2}\tilde{r}_i^2 + \sum_{i<j}^{N}\frac{1}{|\tilde{\boldsymbol{r}}_i - \tilde{\boldsymbol{r}}_j|} \tag{S7}$$

The harmonic well produces a finite, inhomogeneous cluster. Without a neutralizing background, its classical energy is the positive confinement-plus-repulsion potential $U$ in Eq. (S7), whose minimum depends on particle number, shell occupation, and the boundary. Bulk-lattice studies [S4] motivate triangular local coordination; their Madelung energies are not the energy of this finite trap.

To characterize the competition between Coulomb repulsion and quantum delocalization, we define $\kappa$ as the interaction energy at the oscillator length divided by the oscillator energy scale. With $l_q = \sqrt{\hbar/(\gamma m\omega_\perp)}$, the beam-reference coupling is

$$\kappa = \frac{C}{\hbar\omega_\perp l_q} = \frac{e^2}{4\pi\epsilon_0\gamma\,\hbar\omega_\perp}\sqrt{\frac{\gamma m\omega_\perp}{\hbar}}. \tag{S8}$$

Using $\frac{e^2}{4\pi\epsilon_0} = \alpha\hbar c$ gives a form convenient for beam parameters,

$$\kappa = \alpha\sqrt{\frac{mc^2}{\gamma\,\hbar\omega_\perp}}, \tag{S9}$$

where $\alpha = e^2/(4\pi\epsilon_0\hbar c)$ is the fine-structure constant. In the coplanar, laboratory-time Larmor reference, the transverse mass $\gamma m$ and Coulomb coefficient $C = e^2/(4\pi\epsilon_0\gamma)$ yield an effective Bohr radius $\hbar^2/(\gamma m C) = a_0$, so that $\kappa = l_q/a_0$ and the explicit beam-energy dependence cancels from $\kappa(B)$ [Eq. (2) of the main text]. For a given sheet population $N$ across reference focusing fields $B$, we obtain the single-axis root-mean-square (rms) transverse cluster width $\sigma_x(B) = \sqrt{\langle r^2\rangle/2}$ (with $\langle r^2\rangle$ the mean-square radial distance from the trap center) from the dimensionless radial factor $f(B) = \sqrt{\langle r^2\rangle/(2l_q^2)}$ through

$$\sigma_x(B) = f(B)\,l_q(B) \propto [\kappa(B)]^{1/3}\,l_q(B).$$

Here the $\kappa^{1/3}$ scaling reflects the asymptotic strong-coupling balance between harmonic confinement and Coulomb repulsion at fixed $N$ (since $l_{\rm cl}/l_q = \kappa^{1/3}$). Using FN-DMC with forward-walking projection $L = 0.5\,\omega_\perp^{-1}$ ($\Delta\tau = 0.0025\,\omega_\perp^{-1}$, $W = 256$ walkers, four replicas) and the 1 T trial parameters $(\alpha_{\rm orb}, b) = (0.7, 0.64)$, simulations at $B = 0.5$, 0.3, and 0.1 T ($\kappa \simeq 970$, $1.25 \times 10^3$, and $2.17 \times 10^3$) yield finite-projection radial factors $f(B) \simeq 21.2$, 23.1, and 27.8 (carrying a $\sim$ 1% residual finite-projection systematic upward shift in $\sigma_x$ relative to virial and $\kappa^{1/3}$ scaling bounds when $(\alpha_{\rm orb}, b)$ is held fixed across $\kappa$), consistent with $\kappa^{1/3}$ expansion across the sub-Tesla regime. Connecting this effective immersed Larmor trap to a full beamline experiment requires incorporating the axial magnetic lens profile $B_z(z)$, canonical angular momentum, and transverse state preparation.

**Stochastic Gradient Relaxation.** We obtain the classical configurations by numerically relaxing the potential $U$ in Eq. (S7). Decreasing displacement noise assists exploration of configuration space. The update equation is:

$$\mathbf{u}_i^{(k+1)} = \mathbf{u}_i^{(k)} - \Delta s\, \nabla_{\mathbf{u}_i} U + \Delta s\, \sigma_k \boldsymbol{\eta}_i^{(k)}, \qquad (S10)$$

Here $\boldsymbol{u}_i^{(k)}$ denotes the dimensionless transverse position $\tilde{\boldsymbol{r}}_i = \boldsymbol{r}_{\perp,i}/l_0$ of electron $i$ at iteration $k$, $\Delta s$ is a dimensionless pseudo-time step, $-\nabla_{\boldsymbol{u}_i} U$ is the deterministic dimensionless force (the sum of the inward harmonic restoring force $-\boldsymbol{u}_i$ and pairwise Coulomb repulsions), $\boldsymbol{\eta}_i^{(k)}$ is an independent two-dimensional standard normal random vector, and $\sigma_k$ is a decreasing numerical noise amplitude. Numerically, Eq. (S10) implements a simulated-annealing quench: starting from a Gaussian random cloud [Fig. 2(c); $N = 120$, initial single-axis rms width $2l_0$, $\Delta s = 0.01$ over $10^4$ steps with $\sigma_k$ decreasing linearly from 0.5 to 0 over the first $6 \times 10^3$ steps], gradient descent drives configurations toward local mechanical equilibrium while crossing shallow barriers to reduce fivefold and sevenfold disclination defects. In Fig. 2(d) (93 bounded Voronoi cells colored by coordination $z = 5,6,7$), Delaunay bonds within 1.8 times the median nearest-neighbor spacing over the interior $r_i \leq 0.75 r_{\max}$ (65 charges) yield 94% sixfold coordination and a mean local bond-orientational modulus $\langle|\psi_{6,i}|\rangle \simeq 0.82$. Crucially, the accumulated quench parameter $s = k\Delta s$ and iteration count $k$ represent a fictitious numerical annealing schedule used solely to identify static low-energy configurations; they do not represent real-time dissipative cooling or the physical magnetic-focusing transit time of the electron beam (whose physical space-charge rearrangement scale $\tau_C$ is estimated separately below).

**Variational and Diffusion Monte Carlo Methods.** Finite clusters with $N \gtrsim 20$ can exhibit concentric shells and locally sixfold-coordinated cores [S5,S6]. We use VMC and FN-DMC to benchmark the $T_\perp = 0$, fully spin-polarized fixed-node ground state at the representative population $N = 55$ within $N \sim 10\text{–}10^2$. This population fills the lowest $M = 10$ two-dimensional harmonic-oscillator shells [$N = M(M+1)/2$], removing open-shell degeneracy from the noninteracting reference. Measuring transverse coordinates in units of the quantum harmonic-oscillator length $l_q = \sqrt{\hbar/(\gamma m \omega_\perp)}$ ($\boldsymbol{x}_i = \boldsymbol{r}_{\perp,i}/l_q$; labeled $l_0$ on the axes of Fig. 3) connects the quantum and classical dimensionless coordinates through $l_{\rm cl}/l_q = \kappa^{1/3}$ ($\boldsymbol{x}_i = \kappa^{1/3}\tilde{\boldsymbol{r}}_i$). The ratio $\kappa^{1/3}$ compares the classical confinement–repulsion balance length $l_{\rm cl}$ with the oscillator length $l_q$: at $B = 1\,\mathrm{T}$ ($\kappa_0 \simeq 686$), $l_{\rm cl}/l_q \simeq 8.82$ (increasing to 9.9–12.9 across $B =$ 0.5–0.1 T, deep in the Wigner-molecule regime with $r_s = a_{\rm eq}/a_0 \sim 3.5 \times 10^3$–$1.6 \times 10^4$ and harmonic zero-point Lindemann ratio $\lesssim 0.10$), quantifying the separation of these characteristic scales. In the co-moving Larmor reference frame (where the uniform axial magnetic field enters through the restoring frequency $\omega_\perp$), the dimensionless transverse Hamiltonian takes a single-coupling form at fixed particle number and spin polarization:

$$\frac{\mathcal{H}_\perp}{\hbar\omega_\perp} = \sum_i \left[-\frac{1}{2}\nabla_{x_i}^2 + \frac{1}{2}x_i^2\right] + \kappa \sum_{i<j} \frac{1}{|\boldsymbol{x}_i - \boldsymbol{x}_j|}.$$

The trial state combines a complete-shell oscillator determinant of scale $\alpha_{\rm orb}$, a positive same-spin pair factor, and an exact center-of-mass Gaussian correction:

$$\Psi_T = D_{\alpha_{\rm orb}} \exp\left[\sum_{i<j} \frac{(\kappa/3)x_{ij}}{1 + b x_{ij}}\right] \exp\left[-\frac{(1-\alpha_{\rm orb})N|\boldsymbol{X}_{\rm cm}|^2}{2}\right].$$

Our quantum calculation proceeds in three steps based on this ansatz. First, $D_{\alpha_{\rm orb}}$ is the antisymmetric Slater determinant formed by the lowest 10 closed two-dimensional harmonic-oscillator shells with dimensionless inverse-width-squared parameter $\alpha_{\rm orb}$ (scaling orbital coordinates as $\sqrt{\alpha_{\rm orb}}\,\boldsymbol{x}_i$), while the Gaussian factor in $\boldsymbol{X}_{\rm cm} = N^{-1}\sum_i \boldsymbol{x}_i$ keeps the harmonic center-of-mass (Kohn) mode at its exact ground-state width; optimizing $\alpha_{\rm orb} < 1$ at zero pair-Jastrow factor ($u = 0$) expands the internal relative coordinates against space-charge repulsion while coupling particle coordinates solely through the separable center-of-mass mode, defining the Jastrow-free scaled-Slater upper bound $E_{\rm ub}$ [black dashed curve in Fig. 3(d)]. Second, in the VMC stage, multiplying by the pair Jastrow factor $\exp[\sum_{i<j} u(x_{ij})]$ (where $x_{ij} = |\boldsymbol{x}_i - \boldsymbol{x}_j|$ is the dimensionless pair separation and $b$ is the variational Jastrow range parameter) introduces a Coulomb correlation hole by suppressing the relative weight of configurations as $x_{ij} \to 0$, with linear slope $\kappa/3$ enforcing the two-dimensional same-spin Coulomb cusp condition. Sampling $|\Psi_T(\boldsymbol{X})|^2$ via Metropolis random walks selects $(\alpha_{\rm orb}, b)$ to lower the variational energy $\langle\Psi_T|\mathcal{H}_\perp|\Psi_T\rangle/\langle\Psi_T|\Psi_T\rangle$ [using $(\alpha_{\rm orb}, b) = (0.7, 0.64)$ at $B = 1\,\mathrm{T}$ and across the sub-Tesla runs, and coupling-adapted $(\alpha_{\rm orb}, b)$ across $\kappa = 0$–$100$].

Third, FN-DMC [S7] projects the trial state within its fixed nodal surface by evolving $W$ many-body configurations (walkers, each $\boldsymbol{X} = (\boldsymbol{x}_1, \ldots, \boldsymbol{x}_N)$ representing an $N$-electron configuration in $2N$-dimensional space) in dimensionless imaginary time $\tau$ (in units of $\omega_\perp^{-1}$). The imaginary-time propagator $\exp(-\tau\mathcal{H}_\perp/\hbar\omega_\perp)$ suppresses excited-state components; in the importance-sampled mixed distribution $f(\boldsymbol{X},\tau) = \Phi(\boldsymbol{X},\tau)\Psi_T(\boldsymbol{X})$, the projected wave function $\Phi(\boldsymbol{X},\tau)$ approaches the lowest-energy state compatible with the nodal surface of $\Psi_T$. This mathematical projection does not describe physical cooling. With quantum drift velocity $\boldsymbol{v}_D(\boldsymbol{X}) = \nabla_{\boldsymbol{X}}\ln|\Psi_T(\boldsymbol{X})|$, the evolution of $f(\boldsymbol{X},\tau)$ obeys the generalized Fokker–Planck equation

$$\frac{\partial f(\mathbf{X},\tau)}{\partial\tau} = \frac{1}{2}\nabla^2 f(\mathbf{X},\tau) - \nabla\cdot[\mathbf{v}_D(\mathbf{X})f(\mathbf{X},\tau)] - [E_L(\mathbf{X}) - E_T]f(\mathbf{X},\tau) \qquad (S11)$$

where $E_L(\boldsymbol{X}) = \Psi_T(\boldsymbol{X})^{-1}(\mathcal{H}_\perp/\hbar\omega_\perp)\Psi_T(\boldsymbol{X})$ is the dimensionless local energy and $E_T$ is the reference trial energy. Operationally, Eq. (S11) evolves each walker $\boldsymbol{X}$ over each imaginary-time step $\Delta\tau$ by combining random-walk diffusion $[\frac{1}{2}\nabla_X^2 f$, representing quantum zero-point kinetic fluctuations], deterministic drift $[-\nabla_X\cdot(f\boldsymbol{v}_D)$, guiding walkers toward regions of large $|\Psi_T|$ and away from nodal surfaces], and local-energy branching $[-(E_L(\boldsymbol{X}) - E_T)f$, which duplicates walkers in low-local-energy configurations $E_L(\boldsymbol{X}) < E_T$ and eliminates those in high-local-energy configurations $E_L(\boldsymbol{X}) > E_T$], while the fixed-node boundary condition $\Psi_T(\boldsymbol{X}) = 0$ forbids nodal crossings so $f(\boldsymbol{X},\tau)$ remains non-negative. We implement this imaginary-time evolution with drift-limited importance-sampled Monte Carlo walks and constant-population branching reconfiguration ( $\Delta\tau = 0.00125\,\omega_\perp^{-1}$ , $W = 512\text{–}1024$ at 1 T ; $\Delta\tau = 0.0025\,\omega_\perp^{-1}$ , $W = 256$ across $\kappa = 0\text{–}100$ and $B = 0.5\text{–}0.1$ T).

In Fig. 3(d), comparing the fixed-node energy per electron $E_{\mathrm{DMC}}/N$ (blue circles, plotted across $\kappa = 1\text{–}686$ with statistical error bars smaller than the symbol size) against the Jastrow-free scaled-Slater upper bound $E_{\mathrm{ub}}/N$ (black dashed curve) defines the per-electron energy lowering relative to this reference, $\Delta E_{\mathrm{corr}}/N \equiv (E_{\mathrm{ub}} - E_{\mathrm{DMC}})/N \geq 0$ [Fig. 3(d), inset]. Because $E_{\mathrm{ub}}/N$ includes zero-point confinement, Pauli exchange, exact center-of-mass confinement, and internal radial expansion, $\Delta E_{\mathrm{corr}}/N$ measures the additional interparticle correlation energy lowering within the chosen nodal sector: it grows by nearly three orders of magnitude from $0.037\,\hbar\omega_\perp$ at $\kappa = 1$ to $34.27\,\hbar\omega_\perp$ at $\kappa_0 \simeq 686$ ( $B = 1$ T , reaching 43.60–75.51 $\hbar\omega_\perp$ across $B = 0.5\text{–}0.1$ T ), approaching the asymptotic strong-coupling scaling $\Delta E_{\mathrm{corr}}/N \propto \kappa^{2/3}$ (dashed guide in the inset).

The noninteracting limit provides an analytical check; FN-DMC then supplies fixed-node energy benchmarks at finite coupling. In the noninteracting limit ($\kappa = 0$), the $M = 10$ closed-shell reference has an exact ground-state energy $E_0/N = (2M+1)/3 = 7\,\hbar\omega_\perp$ and trap-center mean-square radius $\langle r^2\rangle = 7\,l_q^2$, which broadened-orbital control runs ($\alpha_{\mathrm{orb}} = 0.90$) recover within statistical uncertainty ( $E_{\mathrm{DMC}}/N = 7.000 \pm 0.001\,\hbar\omega_\perp$ and forward-walked $\langle r^2\rangle = 7.01 \pm 0.01\,l_q^2$ at projection intervals $L = 2\text{–}4\,\omega_\perp^{-1}$). In the strong-coupling regime, FN-DMC yields $E_{\mathrm{DMC}}/N = 1059.54 \pm 0.01\,\hbar\omega_\perp$ at the $B = 1$ T reference ($\kappa_0 \simeq 686$, below the Jastrow-free upper bound $E_{\mathrm{ub}}/N = 1093.81\,\hbar\omega_\perp$), and $1334.50 \pm 0.01$, $1581.94 \pm 0.01$, and $2280.95 \pm 0.02\,\hbar\omega_\perp$ across four independent replicas at $B = 0.5$, 0.3, and 0.1 T (compared with $E_{\mathrm{ub}}/N =$ $1378.09$, $1633.90$, and $2356.47\,\hbar\omega_\perp$; quoted uncertainties denote between-replica statistical standard errors at the specified $(\Delta\tau, W)$ protocol).

**Density, structure factor, and conditional density.** Because standard FN-DMC samples the mixed distribution $\Phi_0\Psi_T$, we estimate position-space observables in Figs. 3(a)–3(c) and across sub-Tesla fields using forward-walking descendant weights [S8] over a projection interval $L = 0.5\,\omega_\perp^{-1}$, which suppress mixed-estimator bias toward $|\Phi_0|^2$ while retaining sufficient ancestor diversity at finite walker number (with $n(\boldsymbol{r})$ normalized to $\int n(\boldsymbol{r})\,d^2r = N$). Using the internally aligned sampled configurations defined below, we evaluate the transverse static structure factor as

$$S(\boldsymbol{q}) = \frac{1}{N}\left\langle \left| \sum_{j=1}^{N} e^{i\boldsymbol{q}\cdot\boldsymbol{r}_j} \right|^2 \right\rangle. \quad (S12)$$

It contains both the finite density profile and interparticle correlations, with $S(\boldsymbol{0}) = N$, while the logarithmic mapping $\ln[1 + S(\boldsymbol{q})]$ enhances visibility across dynamic range. Because the circular harmonic trap has no preferred orientation, unaligned ensemble averages smear orientational structure into rings; prior to accumulating Figs. 3(a) and 3(b) (from two $W = 512$ production trajectories), each sampled configuration is therefore rigidly rotated about the trap origin by $-\Theta_6$ to align the sixfold bond-orientational phase $\Theta_6$ of its inner core (evaluated within $r_i \le R_{\text{inner}} \simeq 16.8\,l_q$), without translation or radial rescaling. This postprocessing defines an aligned frame and applies no physical pinning force. The smooth outermost shell in Fig. 3(a) after inner-core alignment is consistent with weak angular locking between the outer shell and the inner core (inter-shell rotational decoupling). To resolve angular correlations within the outer ring [Fig. 3(c)], each electron falling within the outermost shell of center radius $r_0 \simeq 34.3\,l_q$ and half-width $\delta \simeq 0.38\,l_q$ (indicator $I_i = 1$ for $r_0 - \delta < |\boldsymbol{r}_i| < r_0 + \delta$ and $I_i = 0$ otherwise) is used in turn as a reference particle: rotating all other $N - 1$ coordinates by $R(-\theta_i)$ (where $\theta_i = \arg(\boldsymbol{r}_i)$ is the polar angle of electron $i$ and $R$ is the two-dimensional rotation matrix) places the reference electron on the $+x$ axis (cyan marker), yielding the co-rotating conditional number density (with $\delta^{(2)}$ the two-dimensional Dirac delta function)

$$n_{\text{cond}}(\boldsymbol{r}) = \frac{\left\langle \sum_i I_i \sum_{j\neq i} \delta^{(2)}\left[\boldsymbol{r} - R(-\theta_i)\boldsymbol{r}_j\right] \right\rangle}{\left\langle \sum_i I_i \right\rangle}. \quad (S13)$$

Normalized to $\int n_{\text{cond}}(\boldsymbol{r})\,d^2r = N - 1$ (in units of $l_q^{-2}$), $n_{\text{cond}}(\boldsymbol{r})$ probes intrinsic two-body correlations without introducing any physical pinning force: concentric rings reflect radial shell occupation, angular lobes indicate preferred relative particle positions [S9,S10], and the depletion hole around the $+x$ reference marker reflects the combined effects of Pauli exchange and Coulomb repulsion (the white guides in Figs. 3a,b are schematic geometric references). To quantify interior hexagonal ordering across coupling strength, we evaluate the sixfold bond-orientational order parameter $|\psi_6| = \langle | N_{\text{bulk}}^{-1} \sum_{i\in\text{bulk}} n_i^{-1} \sum_{j\in\mathcal{N}_i} e^{i6\theta_{ij}} | \rangle$ over the interior shells ($r_i \le 24.0\, l_q$ at 1 T, scaled across couplings in proportion to the Jastrow-free scaled-Slater rms trap radius—giving $2.35\, l_q$ and $3.11\, l_q$ at $\kappa = 0$ and 1, and approaching $\propto \kappa^{1/3}$ at strong coupling—with Delaunay bonds truncated above 1.8 times the median nearest-neighbor spacing, where $N_{\text{bulk}}$ is the interior electron count and $\mathcal{N}_i$ denotes the $n_i$ Delaunay neighbors of electron $i$ at bond angles $\theta_{ij}$). From the weakly interacting gas ($\kappa = 0$–$1$) to $B = 1$ T ($\kappa_0 \simeq 686$) and into the sub-Tesla regime ($B = 0.5$, 0.3, and 0.1 T), $|\psi_6|$ rises from 0.12 to $0.66 \pm 0.01$ (four-replica $W = 1024$ benchmark; $0.65 \pm 0.01$ for the two $W = 512$ trajectories in Fig. 3) and further to $0.69 \pm 0.01$, $0.68 \pm 0.02$, and $0.74 \pm 0.01$. After subtracting the per-replica coherent mean-density contribution through the connected structure factor $S_{\text{conn}}(\boldsymbol{q}) = S(\boldsymbol{q}) - |\langle \sum_j e^{i\boldsymbol{q}\cdot\boldsymbol{r}_j} \rangle|^2/N$, the replica-averaged normalized sixth-harmonic angular modulation on the primary reciprocal ring increases from 0.03–0.04 ($\kappa \le 1$) to $0.26 \pm 0.01$ (1 T) and 0.34–0.39 (0.5–0.1 T, with between-replica uncertainties $\pm 0.02$–$0.10$), supporting pronounced sixfold correlations in this finite, spin-polarized cluster within the fixed-node approximation.

**Sub-Tesla Larmor-Trap Scales and Phase-Space Preparation Estimates.** Extending the 1 T reference benchmark ($\kappa_0 \simeq 686$, $l_q \simeq 36.3$ nm; Fig. 3) into the sub-Tesla regime ($B = 0.5$, 0.3, and 0.1 T) illustrates how softer magnetic focusing enlarges the spatial sheet while tightening the transverse temperature requirement. Within the effective harmonic Larmor model for an illustrative 200 keV beam ($\gamma \simeq 1.391$), reducing the axial reference field across $B = 0.5$, 0.3, and 0.1 T lowers the Larmor trapping frequency $\omega_\perp = eB/(2\gamma m)$ from $3.16 \times 10^{10}$ to $1.90 \times 10^{10}$ and $0.632 \times 10^{10}$ rad/s ($\hbar\omega_\perp/k_B \simeq 0.241$, 0.145, and 0.048 K), while expanding the quantum oscillator length $l_q = \sqrt{2\hbar/(eB)}$ from 51.3 to 66.2 and 115 nm and increasing the $N$-independent coupling parameter $\kappa(B) = l_q/a_0$ from 970 to $1.25 \times 10^3$ and $2.17 \times 10^3$. Multiplying the finite-projection ($L = 0.5\, \omega_\perp^{-1}$) FN-DMC radial factors $f(B) \simeq 21.2$, 23.1, and 27.8 (for the illustrative $N = 55$ closed-shell reference, with $\sim 1\%$ finite-projection systematic uncertainty) by $l_q(B)$ yields model rms beam waists $\sigma_x = f(B) l_q \simeq 1.09$, 1.53, and 3.19 $\mu$m, effective Wigner–Seitz radii $a_{\text{eq}} = 2\sigma_x/\sqrt{N} \simeq 0.29$, 0.41, and 0.86 $\mu$m [average areal densities $\overline{n} = (\pi a_{\text{eq}}^2)^{-1} \simeq 3.7$, 1.9, and 0.43 $\mu\text{m}^{-2}$], and characteristic Coulomb energies $E_C = C/a_{\text{eq}} \simeq 3.5$, 2.5, and 1.2 meV. For sheet populations $N \sim 10$–$10^2$, a continuum electrostatic-disk estimate in the harmonic trap gives $\sigma_x \propto N^{1/3} l_{\text{cl}}$, so that $a_{\text{eq}} \propto N^{-1/6} l_{\text{cl}}$ and $E_C \propto N^{1/6}$ vary by less than 15% over a twofold change in $N$ [and by $(6\sqrt{N})^{-1} \approx 2\%$ under shot-to-shot Poisson fluctuations $\Delta N \sim \sqrt{N}$ across $N \sim 50$–$100$], indicating weak electrostatic size sensitivity to $N$. Physically, reducing $B$ expands the transverse sheet and relaxes the rms attosecond bunch duration required for a 10% thin-sheet aspect ratio ($\gamma v_0 \tau_b / a_{\text{eq}} = 0.1$) to $\tau_b \simeq 100$, 140, and 300 as, while lengthening the characteristic Larmor-frame space-charge rearrangement time scale $\tau_C = \sqrt{\gamma m a_{\text{eq}}^3 / C}$ from 14 to 23 and 70 ps (corresponding to laboratory propagation lengths $v_0 \tau_C \simeq$ 2.9, 4.8, and 15 mm, a physical dynamical scale distinct from the numerical quench steps in Eq. (S10); unconfined axial expansion doubles $\gamma\Delta z$ from $0.1\,a_{\mathrm{eq}}$ within $\sqrt{0.1}\,\tau_C \simeq 0.32\,\tau_C$ independently of $N$ and $B$, motivating spatial preconditioning prior to peak compression).

Adopting the classical strong-coupling target $\Gamma = E_C/(k_B T_\perp) \geq 100$ sets the model transverse-temperature budget in the Larmor frame to $T_\perp \lesssim 0.41, 0.29,$ and $0.14$ K at $B = 0.5, 0.3,$ and $0.1$ T, corresponding to an intrinsic normalized thermal transverse emittance budget $\varepsilon_n = \sigma_x\sqrt{\gamma k_B T_\perp/(mc^2)} \lesssim 11$, $13$, and $18$ pm rad (distinct from the coherent rigid-body Larmor rotation removed in the rotating frame [S11]; reaching the crystallization threshold $\Gamma_m \simeq 130\text{–}140$ [S6] tightens $T_\perp$ by $\sim 30\%$ and $\varepsilon_n$ by $\sim 15\%$ ). At the cathode level, where $\varepsilon_{n,\mathrm{src}} = \sigma_{x,\mathrm{src}}\sqrt{\mathrm{MTE}/(mc^2)}$, a cryogenic single-crystal $\mathrm{Cu}(100)$ photocathode with mean transverse energy $\mathrm{MTE} \simeq 5$ meV [S12] gated to an effective emission radius $\sigma_{x,\mathrm{src}} \sim 110\text{–}180$ nm provides a compatible source thermal-emittance scale prior to beamline transport. Realizing this ordered regime in a propagating beam therefore requires combining low-emittance cathode preparation with spatial preconditioning to mitigate disorder-induced heating [S13], synchronized longitudinal compression, and controlled focal transport.

Address correspondence to Y.P. (†yiming.pan@shanghaitech.edu.cn) and H.G. (††genghao@nuaa.edu.cn).